\documentclass[prl,aps,twocolumn,showpacs,floatfix]{revtex4}
\usepackage{graphicx}
\usepackage{amsmath}

\newcommand{\tcr}{t_{\rm cross}}
\newcommand{\tdel}{t_{\rm delay}}
\newcommand{\tmax}{t_{\rm max}}
\newcommand{\omax}{\Omega_{\rm max}}

\begin{document}

\title{Measurement back-action on adiabatic coherent electron transport}

\smallskip

\author{J\'er\^ome Rech$^{1,2}$ and Stefan Kehrein$^{1}$} 
\affiliation{$^1$ Physics Department, Arnold Sommerfeld Center for Theoretical Physics and Center for NanoScience, \\ 
Ludwig-Maximilians-Universit\"at, Theresienstrasse 37, 80333 Munich, Germany \\
$^2$ Centre de Physique Th\'eorique, UMR 6207, Case 907, Luminy, 13288 Marseille Cedex 9, France}
\date{\today} 
\begin{abstract}

We study the back-action of a nearby measurement device on electrons
undergoing coherent transfer via adiabatic passage (CTAP) in a
triple-well system. The measurement is provided by a quantum point
contact capacitively coupled to the middle well, thus acting as a
detector sensitive to the charge configuration of the triple-well
system. We account for this continuous measurement by treating the
whole $\{$triple-well + detector$\}$ as a closed quantum system. This
leads to a set of coupled differential equations for the density
matrix of the enlarged system which we solve numerically. This
approach allows to study a single realization of the measurement
process while keeping track of the detector output, which is especially
relevant for experiments. In particular, we find the emergence of a
new peak in the distribution of electrons that passed through the
point contact. As one increases the coupling between the middle
potential well and the detector, this feature becomes more prominent
and is accompanied by a substantial drop in the fidelity of the CTAP
scheme.

\end{abstract}
\maketitle

Solid-state based quantum computer architectures are currently the
focus of active experimental and fundamental research, as many of them
offer the promise of being scalable, therefore opening the way to
significant improvements in efficiency of certain algorithms. An
essential feature of many proposals for scalable quantum computation
is the coherent transport over large distances of quantum information,
encoded e.g. in electron spins.

A recent method of all-electrical population transfer for electrons
has been suggested in solid-state systems consisting of a chain of
potential wells \cite{greentree}. Termed Coherent Transfer by
Adiabatic Passage or CTAP, this technique is a spatial analogue of the
STImulated Raman Adiabatic Passage (STIRAP) protocol \cite{stirap}
used in quantum optics to transfer population between two long-lived
atomic or molecular energy levels. The CTAP scheme amounts to
transporting electrons coherently from one end of the chain to the other
by dynamically manipulating the tunnel barriers between the successive
potential wells. For the appropriate driving of the system, this
method ensures that the occupation in the middle of the chain is
exponentially suppressed at any point of time. 

There have been several proposals to perform the CTAP scheme in
 triple-well solid-state systems such as quantum dots
\cite{schoerer,qdot-proposal,Cole2008,Das2009}, 
ionized donors \cite{PinSi-proposal} or
superconductors \cite{sc-proposal}. Similarly, this technique has been
put forward as a means to transport single atoms
\cite{coldatoms-proposal} and Bose-Einstein condensates
\cite{bec-proposal} within optical potentials. Recently, a classic
analogue of CTAP has also been demonstrated experimentally, using
photons in a triple-well optical waveguide
\cite{photons}. The CTAP protocol therefore has both a quantum optics
and a condensed matter version, thus raising interest well beyond the
field of quantum information.

The implementation of the CTAP protocol naturally brings about the
question of its observation. The most striking signature of CTAP is the vanishing
occupation of the middle potential well, which can be monitored using
a sensitive electrometer. In solid-state devices, this is usually
achieved using ballistic quantum point contacts (QPC). The electric
current flowing through the QPC is influenced by the presence of an
electronic charge in its close environment, thus turning the QPC into
a charge detector. While this could provide a good test for the
observation of CTAP, one may wonder to what extent this charge
detection is invasive. This is also directly relevant for possible
applications of the CTAP scheme for quantum information purposes:
measurement back-action in the above setup can be related to 
the influence of a decohering environment along the chain. 

Now it is largely believed that the CTAP scheme is relatively robust
against this type of measurements precisely because the middle
potential well is barely populated. However, recent work
\cite{decoherence} concentrating on the decoherence aspects associated with
nonlocal measurements suggests otherwise. In this Letter, we study the
measurement back-action of the QPC on the CTAP scheme, focusing on the
effects of a continuous measurement process. While the measurement process is 
instantaneous for optical experiments, it can take a significant amount of 
time for typical solid-state setups and therefore leads to such a
continuous evolution of the system subjected to it. The continuous
measurement thus poses a non-trivial time-dependent problem,
complicated by the dynamic tuning of the tunneling rates between
potential wells. Our approach closely follows an alternative
derivation of the Bayesian formalism developed by Korotkov
\cite{korotkov}. The advantages over the conventional master equation are 
two-fold. First, it allows for the analysis of a particular realization 
of the measurement process, rather than capturing the behavior of the 
system averaged over many measurements. Second, it enables us to keep 
track of the detector output over the duration of the transfer, providing 
us with information about the time evolution of the system, relevant to 
upcoming experiments.

We consider a triple-well solid-state system, 
whose Hamiltonian is characterized by the time-dependent tunneling rate
$\Omega_{ij} (t)$ between wells $i$ and $j$, and the energy $\epsilon$ of the potential wells:
\begin{equation}
H_{3w} = \sum_{i=1}^3 \epsilon c_i^{\dagger} c_i + \left( \hbar\Omega_{12}(t) c_1^{\dagger} c_2 + \hbar\Omega_{23}(t) c_2^{\dagger} c_3 + \rm{h.c.} \right) ,
\label{eq-h3w}
\end{equation}
where $c_i^{\dagger}$ creates an electron in well $i$.

We now wish to apply the CTAP scheme to coherently transfer an
electron from well 1 to well 3. Following Ref.~\onlinecite{greentree},
this is achieved by applying Gaussian voltage pulses to tune the
tunnel barriers in time according to
\begin{eqnarray}
\Omega_{12}(t) & = & \omax \exp \left[ - \frac{(t - \tmax/2 - \tdel)^2}{2 \sigma^2} \right] \nonumber \\
\Omega_{23}(t) & = & \omax \exp \left[ - \frac{(t - \tmax/2)^2}{2 \sigma^2} \right] ,
\end{eqnarray}
where we introduced the pulses height ($\omax$) and duration
($\tmax$), and chose for the standard deviation $\sigma = \tmax / 8$. The delay between pulses
is selected in order to optimize the transfer, $\tdel = 2 \sigma$
\cite{delay}. Like in the STIRAP protocol, the pulses are applied in 
a counter-intuitive sequence, where $\Omega_{23}$ is fired prior to
$\Omega_{12}$, in order to maximize the fidelity of the transfer 
\cite{greentree}.

To monitor the charge configuration of this system, we couple
the middle potential well to a charge detector. Here we use a simplified 
version of the QPC, namely a tunnel junction whose barrier height is 
sensitive to the presence of an electron in the middle well. The hopping 
amplitude through the barrier varies from $\Omega'$ to $\Omega$, depending 
on whether or not well 2 is occupied \cite{gurvitz}. The detector Hamiltonian can thus be written as
\begin{eqnarray}
H_{qpc} & = & \sum_{r} E_{r} : a_{r}^{\dagger} a_{r} : + \sum_{l} E_{l} : a_{l}^{\dagger} a_{l} :  \nonumber \\
& & + \sum_{l,r} \hbar \left( \Omega + \delta \Omega~ c_2^\dagger c_2\right) ( a_{r}^{\dagger} a_{l} +  a_{l}^{\dagger} a_{r}) ,
\end{eqnarray}
where $a_{r}^{\dagger}$ and $ a_{l}^{\dagger}$ are the electron
creation operators in the right and left electrode respectively, while
$E_{r,l}$ stands for the set of energy levels in the reservoirs. Here we introduced $\delta \Omega = \Omega' -
\Omega$, and assumed all tunneling amplitudes to be real and 
independent of the states in the electrodes.

Our goal is to study the evolution of the triple-well system under
continuous measurement by the detector, focusing on a single realization of
the measurement process. This is achieved by considering the triple-well 
system and the detector as the two parts of an enlarged quantum system. This allows for
describing the quantum state of this enlarged system via a generalized
density matrix $\rho_{ij}^n$(t) \cite{gurvitz}. The latter corresponds to
the density matrix in the basis of localized states (associated with
wells 1, 2, and 3) further divided into components with different
number $n$ of electrons passed through the detector. The evolution of
this generalized density matrix is given by a set of Bloch-type
equations obtained from the many-body Schr\"odinger equation for the
entire system. Extending the derivation of Ref.~\onlinecite{gurvitz} to include
time-dependent tunneling rates, one obtains the following set of rate
equations for the density matrix
\begin{eqnarray}
\dot{\rho}_{11}^n (t) & = & D \left[\rho_{11}^{n-1}(t) - \rho_{11}^n (t)\right] - 2 \Omega_{12}(t) {\rm Im} \left[\rho_{12}^n (t)\right] \nonumber \\
\dot{\rho}_{22}^n (t) & = & D'  \left[\rho_{22}^{n-1}(t)- \rho_{22}^n (t)\right] + 2 \Omega_{12}(t) {\rm Im} \left[\rho_{12}^n (t)\right] \nonumber \\
& & - 2 \Omega_{23}(t) {\rm Im} \left[\rho_{23}^n (t)\right] \nonumber \\
\dot{\rho}_{33}^n (t) & = & D \left[\rho_{33}^{n-1}(t) - \rho_{33}^n (t)\right] + 2 \Omega_{23}(t) {\rm Im} \left[\rho_{23}^n (t)\right] \nonumber \\
\dot{\rho}_{12}^n (t) & = & \sqrt{D D'} \rho_{12}^{n-1}(t) - \frac{D+D'}{2} \rho_{12}^n (t) \nonumber \\
& & + i \Omega_{12} (t) \left[ \rho_{11}^n (t) - \rho_{22}^n (t)\right] + i \Omega_{23} (t)  \rho_{13}^n (t) \nonumber \\
\dot{\rho}_{23}^n (t) & = & \sqrt{D D'} \rho_{23}^{n-1} (t) - \frac{D+D'}{2} \rho_{23}^n (t) \nonumber \\
& & + i \Omega_{23} (t) \left[ \rho_{22}^n (t) - \rho_{33}^n (t)\right] - i \Omega_{12} (t)  \rho_{13}^n (t) \nonumber \\
\dot{\rho}_{13}^n (t) & = & D \left[ \rho_{13}^{n-1}(t)-\rho_{13}^n (t)\right] + i \Omega_{23} (t) \rho_{12}^n (t) \nonumber \\
& & - i \Omega_{12} (t) \rho_{23}^n (t) ,
\label{eqs-bloch}
\end{eqnarray}
where we used the convention $\rho_{ij}^0(t)=0$, and introduced the
tunneling rates $D=2\pi \hbar \rho_R\rho_L\Omega^2 eV$ and $D'=2\pi
\hbar \rho_R\rho_L \Omega'^2 eV$. Here we considered a QPC under a
voltage bias $V$, with constant density of states $\rho_{R,L}$ in the 
electrodes. Note that by tracing over the detector degrees of
freedom, one recovers the conventional master equation for the
density matrix $\rho_{ij} = \sum_n \rho_{ij}^n$. In this case, the
measurement back-action reduces to a constant dephasing term $\Gamma =
(\sqrt{D}-\sqrt{D'})^2/2$, which only affects the coherences
$\rho_{12}$ and $\rho_{23}$.

The drawback in including the detector into the quantum part of the
setup is that one now needs a way to extract classical information
from it. Following Ref.~\onlinecite{korotkov}, we introduce a
classical pointer which periodically collapses the wavefunction of the
quantum system. The pointer only interacts with the detector at times
$t_k$, forcing it to choose a definite value $n_k$ for the number
$n(t_k)$ of electrons that passed through the QPC. The value of $n_k$
is picked randomly according to the probability distribution $P(n)$
set by the density matrix at the time of the collapse, namely $P(n) =
\rho_{11}^n(t_k) + \rho_{22}^n(t_k) + \rho_{33}^n(t_k)$. Once a
particular $n_k$ has been selected, the density matrix should be
immediately updated
\begin{equation}
\rho_{ij}^n (t_k^+) = \delta_{n,n_k} \frac{\rho_{ij}^{n_k}(t_k^-)}{\rho_{11}^{n_k}(t_k^-)+\rho_{22}^{n_k}(t_k^-)+\rho_{33}^{n_k}(t_k^-)}.
\label{eq-collapse}
\end{equation}
After that, the system evolves according to Eq.~(\ref{eqs-bloch}) until
the next collapse at $t=t_{k+1}$. While it is not clear how frequently the collapse
procedure should occur, we could check explicitly that 
our results are insensitive to the choice of $t_k$, 
provided that $\Delta t_k = t_k-t_{k-1}\ll\tmax$.

The combination of the coupled Bloch equations (\ref{eqs-bloch}) with
the collapse procedure (\ref{eq-collapse}) makes the problem difficult
to solve analytically. Instead, we propose a solution by stochastic
sampling, a method designed to mimic the experimental setup. For a
given sequence of times $t_k$, we numerically implement the succession
of evolutions and collapses while keeping track of the detector
output, i.e. the number of electrons $n(t)$ that passed through the
QPC. This provides us with one particular realization of the measurement
process. In particular, this implies that by the end of the
``simulated experiment'' at $t=\tau$, one obtains a definite answer
whether the electron sits in the middle well or not, so that
$\rho_{22}^{n(\tau)}(\tau)$ can only take the values 0 or 1. Moreover,
reproducing this procedure over thousands of realizations allows us to
extract statistical properties, such as the distribution ${\cal P}(n)$
of the total number of electrons that crossed the tunnel junction over
the duration $\tau$ of the experiment. Such features could be
qualitatively compared to experimental data when the latter becomes
available.

\begin{figure}[tbp]
\centering
 \resizebox{.43\textwidth}{!}{\includegraphics*{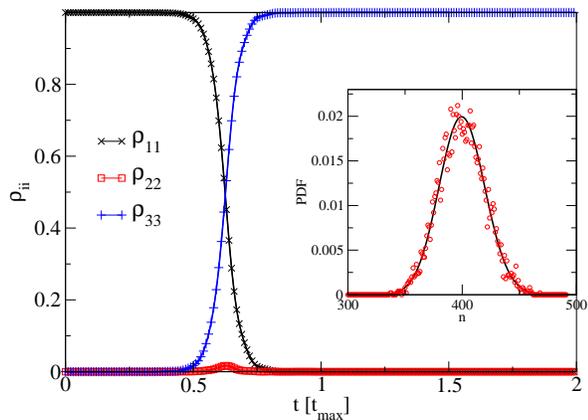}} 
\caption{
Comparison between $\rho_{ii}(t)$
(lines) and $\rho_{ii}^{n_k}(t_k)$ (symbols), for $\omax=80$ and
$D=D'=200$ (in units of $\tmax^{-1}$) and $\tau=2\tmax$. Inset: 
Distribution ${\cal P}(n)$ of the total number of electrons that
crossed the detector over the duration of the experiment,
obtained for 5000 runs (symbols). The black curve corresponds to 
the expected Poisson distribution}
\label{fig-decoupled}
\end{figure} 

As a first test of the method, we consider the case of a
decoupled detector, obtained by setting equal tunneling
rates $D=D'$ in Eq.~(\ref{eqs-bloch}). In this case, no information on
the position of the electron can be extracted from the measurement and
the CTAP scheme works with optimum fidelity. Turning to the detector
output, we plot as an inset in Fig.~\ref{fig-decoupled} the distribution ${\cal
P}(n)$ obtained after a few thousand runs, which turns out to be very
close to the expected Poisson behavior. Furthermore, the profile of
the diagonal density matrix $\rho_{ii}^{n_k}(t_k^+)$ is identical to
the one obtained from solving the conventional master equation for
$\rho_{ii}(t)$ with Hamiltonian $H_{3w}$ (\ref{eq-h3w}), see
Fig.~\ref{fig-decoupled}.

\begin{figure}[tbp]
\centering
 \resizebox{.43\textwidth}{!}{{\Large (a)}\includegraphics*{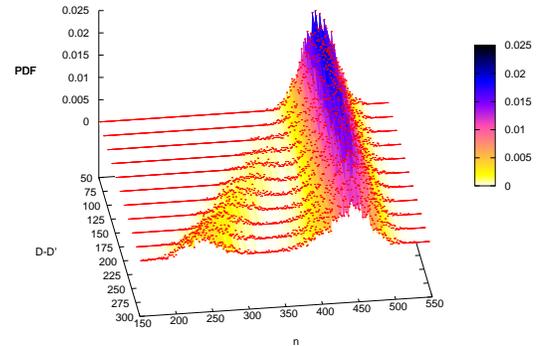}}
 \resizebox{.43\textwidth}{!}{{\Large (b)}\includegraphics*{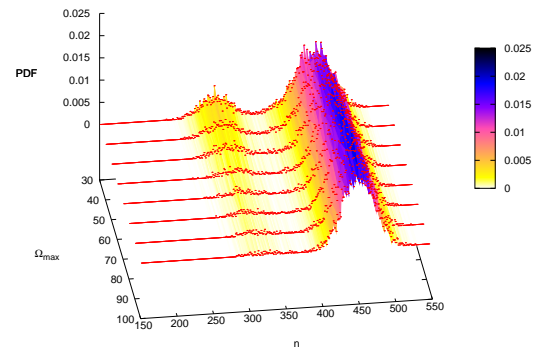}}
 \caption{
Distribution ${\cal P}(n)$ as a function of (a) $D-D'$ with
$\omax=50$, $D=300$ (in units of $\tmax^{-1}$), (b) $\omax$, for
$D=300$, $D'=100$ (in units of $\tmax^{-1}$). In both cases, the
duration of the experiment is $\tau=1.5 \tmax$.}
\label{fig-distribution}
\end{figure} 

\begin{figure}[tbp]
\centering
 \resizebox{.43\textwidth}{!}{\includegraphics*{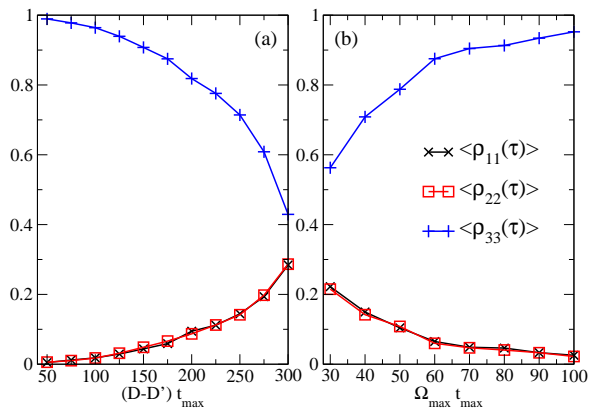}}
\caption{
Diagonal part of the density matrix $\rho_{ii}^{n(\tau)}(\tau)$
averaged over all runs, as a function of (a) $D-D'$ with $\omax=50$,
$D=300$ (in units of $\tmax^{-1}$), (b) $\omax$, for $D=300$, $D'=100$
(in units of $\tmax^{-1}$). In both cases, the duration of the
experiment is $\tau=1.5 \tmax$.}
\label{fig-fidelity}
\end{figure}

Let us now increase the coupling to the detector, by reducing the 
value of the tunneling rate $D'$ compared to $D$. 
Keeping track of the distribution ${\cal P}(n)$ of
the number of electrons in the detector as a function of the rate
mismatch $D-D'$ leads to the results of Fig.~\ref{fig-distribution}a. 
The main signature of the measurement back-action on this distribution
is the emergence of a satellite peak on top of the
Poisson-like behavior. As one increases the value of $D-D'$, making
the measurement stronger, this secondary structure becomes more
prominent, while the main peak flattens out. The distance between
these two features grows like $(D-D')\tau$. Note that while the peaks are 
well-separated for longer times $\tau$, they are also more spread out. 
Out of the
thousands of realizations that result in the distribution of
Fig.~\ref{fig-distribution}a, the ones that contribute to the
satellite peak correspond to situations for which the electron sits in
the middle well by the end of the experiment,
i.e. $\rho_{22}^{n(\tau)}(\tau)=1$. As a result, the integral under
this secondary peak measures the proportion $p_2$ of runs where the
electron is detected in well 2. Obviously, these realizations
exemplify an unsuccessful transfer, and one would thus expect a
reduced fidelity for the CTAP scheme. Indeed, as the coupling to the
detector becomes more important, one obtains more information on the
location of the electron, and the fidelity of the CTAP protocol
decreases, as shown in Fig.~\ref{fig-fidelity}a. This reduction turns
out to be more important than one would anticipate because the average
probability $p_1$ of finding the electron in the first potential well
is not only finite but also increases along with $D-D'$ in a way that
$p_1 \simeq p_2$, see Fig.~\ref{fig-fidelity}a. The measurement
therefore leads to an increased population of the middle well,
compared to the unmonitored CTAP scheme. This effect is qualitatively
similar to that of pure dephasing \cite{greentree}.

We could further check that for a given value of $D-D'$ increasing the
amplitude $\omax$ of the pulses restores the fidelity of the CTAP
protocol, as illustrated in Fig.~\ref{fig-fidelity}b. This however
goes with a significant loss of weight of the satellite peak in the
probability distribution ${\cal P}(n)$ 
(see Fig.~\ref{fig-distribution}b) corresponding to a loss of
information on the location of the electron. These results also confirm
that the position of the satellite peak is independent of $\omax$.

\begin{figure}[tbp]
\centering
 \resizebox{.43\textwidth}{!}{\includegraphics*{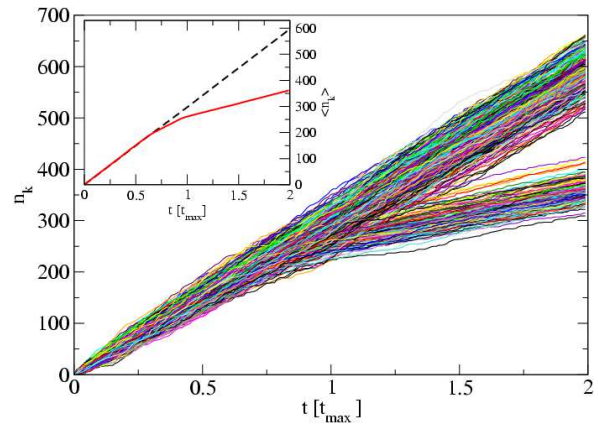}}
\caption{ 
Compiled (main graph) and averaged (inset) record for the upper and lower subsets,
obtained for 1000 runs, with with $\omax=50$, $D=300$, $D'=100$ (in
units of $\tmax^{-1}$) and a duration of the experiment is $\tau=2
\tmax$.}
\label{fig-noft}
\end{figure}

The compiled record of the detector output $n(t)$ for 1000 runs is
plotted in Fig.~\ref{fig-noft}, and one readily sees that it can 
be divided into two subsets. The lower subset is associated
with the ensemble of runs that contribute to the satellite peak in
${\cal P}(n)$, and correspond to the detection of the electron in the
middle well, $\rho_{22}^{n(\tau)}(\tau)=1$. The upper subset is
associated with the ensemble of runs contributing to the main
Poisson-like structure in ${\cal P}(n)$, and correspond to
$\rho_{22}^{n(\tau)}(\tau)=0$. It is instructive to evaluate the average behavior of
$n(t)$ within each of these subsets (see inset of Fig.~\ref{fig-noft}). On
average, for the upper subset one has $\langle n(t) \rangle \sim D t$ 
over the whole duration of the experiment. 
For the lower subset, however, there are three distinct regimes, 
defined by the typical scales $\tmax$ and $\tcr$
(defined as $\Omega_{12}(\tcr)=\Omega_{23}(\tcr)$). For $0 \leq t \lesssim \tcr$, the average  number
$\langle n(t) \rangle$ of electrons through the detector 
 is again given by $D t$ and the average occupation of the middle
well stays much lower than 1. For times $t \gtrsim \tmax$, $\langle n(t) \rangle$
grows like $D' t$ and the average
occupation of the middle well stays close to 1, over the whole time
range: the electron has been detected. The most interesting behavior
occurs for $\tcr \lesssim t \lesssim \tmax$: the detection builds up as the average occupation of
the middle well rises from nearly 0 to nearly 1. In the meantime,
 the average detector output $\langle n(t)
\rangle$ grows like $(D+D')t/2$, while one might have expected a
gradual decrease of the slope from $D$ to $D'$.

Our results can easily be extended to other pulse shapes and sequences
as well as more complicated setups. Our approach also
offers the possibility to reconstruct the density matrix from a given
experimental measurement record of $n(t)$, by replacing the random
collapse procedure with the provided record from the experiment

In summary, we have proposed an approach which allows to study the
measurement back-action on an electron submitted to the CTAP protocol
in a triple-well system. Our work captures the loss of fidelity of the
CTAP scheme associated with the measurement process, a feature
generally accounted for in the conventional master equation formalism
by explicitly adding a dephasing term. The key observation of this Letter 
is that the reduction of the fidelity
is directly connected to the amount of information one can extract
concerning the location of the electron. This has the important implication
that a decohering environment coupling to an electron on the CTAP chain reduces the fidelity
of this scheme for quantum information transfer in spite of the fact
that the occupation probability along the chain can be made arbitrarily small
for the case without decoherence.

We are grateful to J.H. Cole, A. Korotkov and S. Ludwig for helpful discussions.
This work was supported through SFB 631 of the Deutsche Forschungsgemeinschaft, the Center for NanoScience (CeNS) Munich, and the German Excellence Initiative via the Nanosystems Initiative Munich (NIM).

\end{document}